\documentclass[prb,twocolumn,showpacs]{revtex4}
\usepackage{graphicx}
\usepackage{bm}
\usepackage{amssymb,amsmath}
\usepackage[usenames]{color}
\usepackage{epsfig}
\usepackage{hyperref}
\usepackage{subfigure}
\usepackage[sans]{dsfont}
\usepackage{supertabular}
\hypersetup{colorlinks=true, citecolor=blue,
linkcolor=blue,urlcolor=blue }

\begin{document}

\title{ Numerical simulation of the localization of elastic waves in two-
and three-dimensional heterogeneous media}

\author{Reza Sepehrinia\textsuperscript{1}, M. Reza Rahimi Tabar\textsuperscript{1,2}, and Muhammad Sahimi\textsuperscript{3,*}\\
\textsuperscript{1}\textit{Department of Physics, Sharif University of
Technology, Tehran
11155-9161, Iran}\\
\textsuperscript{2}\textit{Institute of Physics, Carl von Ossietzky University, Oldenburg D-26111, Germany}\\
\textsuperscript{3}\textit{Mork Family Department of Chemical
Engineering \& Materials Science, University of Southern
California,\\
Los Angeles, California 90089-1211, USA}}

\begin{abstract}
Localization of elastic waves in two-dimensional (2D) and
three-dimensional (3D) media with random distributions of the Lam\'e
coefficients (the shear and bulk moduli) is studied, using extensive
numerical simulations. We compute the frequency-dependence of the
minimum positive Lyapunov exponent $\gamma$ (the inverse of the
localization length) using the transfer-matrix method, the density
of states utilizing the force-oscillator method, and the
energy-level statistics of the media. The results indicate that all
the states may be localized in the 2D media, up to the disorder
width and the smallest frequencies considered, although the
numerical results also hint at the possibility that there might a
small range of the allowed frequencies over which a mobility edge
might exist. In the 3D media, however, most of the states are
extended, with only a small part of the spectrum in the upper band
tail that contains localized states, even if the Lam\'e coefficients
are randomly distributed. Thus, the 3D heterogeneous media still
possess a mobility edge. If both Lam\'e coefficients vary spatially
in the 3D medium, the localization length $\Lambda$ follows a power
law near the mobility edge, $\Lambda\sim(\Omega-\Omega_c)^{-\nu}$,
where $\Omega_c$ is the critical frequency. The numerical simulation
yields, $\nu \simeq 1.89\pm 0.17$, significantly larger than the
numerical estimate, $\nu\simeq 1.57\pm 0.01$, and $\nu=3/2$, which
was recently derived by a semiclassical theory for the 3D Anderson
model of electron localization. If the shear modulus is constant but
the bulk modulus varies spatially, the plane waves with transverse
polarization propagate without any scattering, leading to a band of
completely extended states, even in the 2D media. At the mobility
edge of such media the localization length follows the same type of
power law as $\Lambda$, but with an exponent, $\nu_T\simeq 1/2$, for
both 2D and 3D media.
\end{abstract}

\pacs{62.30.+d, 05.10.Cc, 71.23.An}

\maketitle

\begin{center}
{\bf I. INTRODUCTION}
\end{center}

Propagation of elastic waves in random media has been a subject of much
interest for several decades. The reason for the interest is at least twofold.
One is that propagation of elastic waves in rock provides much information
on its structure and content.$^{1,2}$ For example, seismic wave propagation and
reflection are used for not only estimating the hydrocarbon contents of an oil
or natural gas reservoir, but also obtaining information on the spatial
distributions of the reservoir's fractures, faults, and strata, as well as its
porosity. They are also the main tool for imaging rock structure over a wide
area, ranging from the Earth's near surface to the deeper crust and upper
mantle. How the inhomogeneities within the Earth's crust affect propagation of
elastic waves has also been, for a long time, a subject of much interest.$^3$
Other rock-related phenomena and problems in which propagation of elastic waves
plays a significant, and often fundamental, role include the analysis of
seismic records for earthquakes in order to develop a theory for predicting
when and where an earthquake may occur,$^4$ and detecting underground nuclear
explosions.

The second reason for the interest in understanding elastic wave propagation
is related to the characterization of materials, the effect of heterogeneities
on their macroscopic properties, and development of a link between their static
and dynamical properties.$^5$ This problem has also been studied for several
decades.$^6$ In particular, propagation of elastic waves has been a major tool
for nondestructive evaluation of composite materials, and gauging the effect
that the defects have on their properties.$^{5,7}$

As elastic waves propagate in a disordered material, the heterogeneities cause
multiple scattering of, and interference in, the waves. The scattering process
modifies both the travel time and amplitudes of the propagating waves. An
important question, then, is whether the heterogeneities and the associated
scattering and interference phenomena can give rise to {\it localization} of
the elastic waves. By localization, we mean a phenomenon in which, over finite
lengths scales (which could, however, be quite large), a wave's amplitude
decays and eventually vanishes. It was recently reported,$^8$ through elegant
experiments, that seismic waves propagating in rock samples exhibit {\it weak}
localization. The fundamental mechanism for weak localization is {\it
constructive interference} of an incident beam, travelling in a given
scattering path, and the waves that move along the same path in the
backscattering direction. Constructive interference exists only in a narrow
range of the angles around the backscattering direction. The typical width of
the cone is of the order of $\zeta/\ell$, where $\zeta$ is the wavelength and
$\ell$ the mean-free path of the waves. Thus, for a medium with a high density
of the scatterers, the backscattering cone is wider. Indeed, the criterion for
the existence of the weak localization regime is, $\zeta/\ell\gg 1$. Weak
localization is also important due to it being a precursor to the strong
localization, which is also the result of multiple scattering by a spatial
distribution of scatterers. As is well known, the main consequence of
strong localization is the absence of diffusion of waves over length scales
that are larger than the localization length.$^9$

Localization of elastic waves in disordered media have important practical
implications. Consider, as an example, propagation of elastic waves in rock.
If the waves do localize, then, their scattering and reflection by the rock
can provide useful information on its structure only over length scales that
are on the order of the localization length of the waves, or smaller. Thus, if,
for example, a station that records information on the seismic waves that
emanate from an earthquake epicenter is farther from the center than the
localization length of the waves, the records cannot provide much useful
information on the phenomena that led to the earthquake and its aftermath.

A very useful quantity for determining whether a state is delocalized or
localized is the minimum positive Lyapunov exponent $\gamma$, which is simply
the inverse of the localization length $\xi$. If $\gamma>0$ for all the
energies or frequencies $\omega$, then, all the states are localized; that is,
the wave function $\psi(r)$ decays at large distances $r$ from the center of
the material's domain as, $\psi(r) \sim \exp[-\gamma(\omega)r]$. Another useful
property is the vibrational density of states of disordered elastic materials,
which depends strongly on the strength of the disorder in the materials.

In the present paper we carry out extensive numerical simulations in order to
study the frequency-dependence of several properties of disordered elastic
media in both two- and three-dimensional (3D) media, and investigate the
conditions under which elastic waves in such media may become localized. We
compute the localization properties of disordered elastic media as a function
of the frequency using the transfer-matrix method. In particular we compute the
frequency-dependence of the minimum positive Lyapunov exponent near the
localization-delocalization transition. The frequency-dependent properties that
we study also include the statistics of the energy levels and the distribution
of the spacings (gaps) between the nearest-neighbor levels. The distribution of
the level spacings for certain matrices has been studied through the theory of
random matrices.$^{10,11}$ It has been shown that the statistics of the energy
levels in the metallic regime of the Anderson localization$^9$ follow the
well-known Wigner-Dyson statistics.$^{10}$ The symmetries of the Hamiltonian
that describe a phenomenon in a disordered medium affect the universality class
of the transition and the distribution function of the level spacings.
Therefore, it should be interesting, as well as important, to study whether
the statistics that we compute for elastic waves in disordered media fall in
any known universality class, such as that of the Anderson model, or that they
give rise to a new class.

To carry out the numerical simulations, we consider a medium with a constant
density and continuous spatial distributions of the Lam\'e coefficients. The
governing equations for the propagation of elastic waves in such a medium are
then discretized and solved. The discretization does, of course, introduce a
cutoff length scale into the problem, namely, the lattice spacing or the linear
size of the blocks in the computational grid. Thus, the results are valid for
the wavelengths that are larger than the basic linear size of the blocks.
At the same time, though, the model that we study is one that has been used
extensively in the geophysics literature for representing the propagation of
seismic waves. Thus, our results are directly relevant to seismic wave
propagation in heterogeneous rock.

In addition to the above considerations, the present study is also motivated
by, and represents a continuation of, our recent study$^{12}$ of propagation of
elastic waves in 2D disordered media. In that study the Martin-Siggia-Rose
method$^{13}$ was used, and the one-loop dynamic renormalization group (RG)
equations were derived for the coupling constants, in the limit of low
frequencies (long wavelengths). The RG analysis made it possible to identify
those regions in the coupling constants space in which the elastic waves are
localized or extended. Thus, using extensive numerical simulations, we aim in
the present paper to check the predictions of the RG analysis carried out
previously.$^{12}$ In addition, our work is relevant to phonon localization in
disordered solids that has been studied extensively in the past. Such a
phenomenon has been studied classically, using both the scalar and vector
models of vibrations in disordered materials,$^{14-16}$ although the vector
models used previously are different from what we consider in the present
paper.

The rest of this paper is organized as follows. In the next section we describe
the model of the heterogeneous elastic media and the governing equations for
elastic wave propagation in such media that we study in this paper. Section III
describes the transfer-matrix computation of the Lyapunov exponents, while the
calculation of the density of states is described in Sec. IV. The results are
presented and discussed in Sec. V, while Sec. VI summarizes the paper.

\begin{center}
{\bf II. MODEL AND GOVERNING EQUATIONS}
\end{center}

Many of the theoretical studies of wave propagation in heterogeneous
media (such as rock) are based on the elastic wave equation
\begin{equation}
m\frac{\partial^2 u_i}{\partial t^2}=\partial_j\sigma_{ij}\;,
\end{equation}
which represents the equation of motion for an elastic medium with mean density
$m$. Here, $u_i$ is the displacement in the $i$th direction, $\sigma_{ij}$
the $ij$th component of the stress tensor $\mbox{\boldmath$\sigma$}$, and $t$
is the time. As usual, $\sigma_{ij}$ is written in terms of the strain tensor,
\begin{equation}
\sigma_{ij}=2\mu({\bf x})u_{ij}+\lambda({\bf x})u_{kk}\delta_{ij}\;,
\end{equation}
where $u_{ij}$ is $ij$ component of the the strain tensor, and $\mu({\bf x})$
and $\lambda({\bf x})$ are the spatially-varying Lam\'e coefficients. For small
deformations of the medium the strain tensor has the following linear form in
terms of the displacement components $u_i$ and $u_j$,
\begin{equation}
u_{ij}=\frac{1}{2}\left(\partial_i u_j+\partial_ju_i\right)\;.
\end{equation}
In the computer simulations that are described below we take the Lam\'e
coefficients $\lambda$ and $\mu$ to be uniformly distributed in the intervals
$[\lambda_0-W_\lambda,\lambda_0+W_\lambda]$, and $[\mu_0-W_\mu,\mu_0+W_\mu]$,
where $\lambda_0$ and $\mu_0$ are the mean values of the coefficients.

\begin{figure}[t]
\epsfxsize8truecm \epsffile{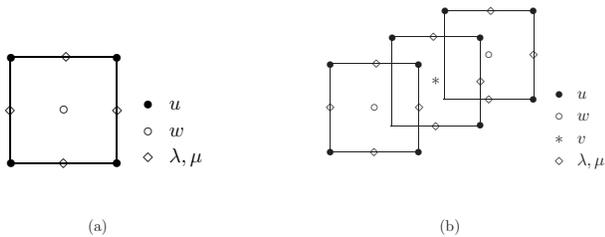} \caption{(a) Two- and (b)
three-dimensional staggered grids used in the simulations. Symbols
on the grid blocks indicate the grid points at which the associated
quantities are evaluated.}
\end{figure}

Accurate numerical simulation of Eq. (1) [together with Eqs. (2) and (3)] is
difficult, particularly if one solves it for a heterogeneous medium in which
the Lam\'e coefficients vary spatially. There have been many attempts in the
geophysics literature to solve Eq. (1) numerically in the time
domain,$^{17-20}$ using a variety of schemes and computational grids.
Typically, in these works,$^{17-20}$ Eq. (1) was discretized by the
finite-difference (FD) technique. However, use of the central-difference FD
method, together with cubic computational grids, can give rise to certain
instabilities in the solution. It has been shown$^{21}$ that stable numerical
solutions are obtained if one uses a staggered computational grid, in order to
define the variables and discretize Eq. (1), unless the medium contains
singularities (such as cracks) or free surfaces, in which case a rotated
staggered grid may be more appropriate.

In the present study we used the staggered computational grids, shown in Fig.
1, in the numerical simulations of Eq. (1). Discretizing Eq. (1) on such a grid
leads to a symmetric Hamiltonian, as expected. We then seek monochromatic
solutions of Eq. (1) for a given frequency, in the form,  $u_i({\bf x},t)=
u_i({\bf x})\exp(i\omega t)$. Writing, ${\bf u}=(u,w)$ in 2D and, ${\bf u}=
(u,w,v)$ in 3D, and discretizing Eq. (1) by the FD method on the staggered grid
shown in Fig. 1, we obtain a set of discretized equations for the determination
of the monochromatic solutions. The resulting discretized equations are given
in the Appendix.

The problem of solving the set of the discretized equations, Eqs. (A1), (A2),
and (A5) - (A7) of the Appendix is then formulated as one of an eigenvalue
problem. If we define a vector {\bf Z}, the components of which represent all
the field variables (displacements) at the grid points, then the set of the
discretized equations is written as
\begin{equation}
\sum_\beta H_{\alpha\beta}Z_\beta=\Omega Z_\alpha\;,
\end{equation}
where, $\Omega=\omega^2$, with the matrix of the coefficients {\bf H} being
symmetric. As mentioned above, discretizing the governing equations introduces
a cutoff length scale and, hence, a cutoff frequency in the simulations, which
do not exist in the continuum equations (1). We neglect such difference between
the discrete and continuous system (which can be reduced by decreasing the size
of the blocks in the computational grid).

\begin{center}
{\bf III. TRANSFER-MATRIX CALCULATIONS}
\end{center}

To determine whether an eigenstate is localized or extended in the
thermodynamic (large system size) limit, we calculate the minimum positive
Lyapunov exponent $\gamma_m$, which is simply the inverse of the localization
length. The most suitable numerical method for directly computing the
localization properties of noninteracting disordered media is, perhaps, the
transfer-matrix (TM) technique, using a strip (bar) in two (three) dimensions,
with periodic boundary conditions in the transverse direction(s). To formulate
the TM computations, we rewrite the difference Eqs. (A1), (A2), and (A5) - (A7)
in the following form
\begin{eqnarray}
\left(\begin{array}{c}
{\bf Z}_{n+1} \\ {\bf Z}_n
\end{array}\right)={\bf T}_n\left(
\begin{array}{c}
{\bf Z}_n \\ {\bf Z}_{n-1}
\end{array}\right)\;,
\end{eqnarray}
where ${\bf Z}_n$ is the vector that contains the values of (the discretized)
displacements ${\bf u}({\bf x})$ in slice number $n$ in the 2D strip or the
3D bar. For 2D media, for example, ${\bf Z}_n$ contains $2M$ components because
every grid point is characterized by two displacements $(u,w)$ (see Fig. 1).
Thus, the vectors on both sides of Eq. (5) contain $4M$ components and, as a
result, ${\bf T}_n$ is a $4M\times 4M$ matrix, resulting in $4M$ Lyapunov
exponents, half ($2M$) of which are independent as they appear in pairs,
$(\gamma,-\gamma)$.

Because the discretized equations are defined on a staggered grid (Fig.
1), and we have a set of coupled equations for each grid points, instead of a
single equation (because we solve a vector equation), the slices for the
TM steps should be defined carefully. In 2D we impose the boundary condition on
one side of the strip and then move forward in the (longitudinal) $x$ direction
by multiplication of the TM matrices. To do this, we define the set of two
lines, $x=i$ and $x=i+\frac{1}{2}$, as a single slice. All the variables in the
slice $(i+1,i+1+\frac{1}{2})$ are then computed, if one knows the values in the
slices $(i,i+\frac{1}{2})$ and $(i-1,i-1+\frac{1}{2})$. In the same way, a
boundary condition is imposed on one end plane of the bar, and a 3D slice is
defined as being composed of two planes, $z=k$ and $z=k+\frac{1}{2}$.

We used $M\times L$ strips and $M\times M\times L$ bars, where $M$ is the
transverse dimension, and $L$ the length. If $N=dM^{d-1}$, where $d=2$ and 3,
then, the dimensions of the TM in Eq. (5) and number of the Lypunov exponents
are $2N$. Therefore, the simulations start with $2N$ initial orthonormal
vectors, corresponding to the dimension of the TM, taken to be, ${\bf v}=
(1,0,\cdots,0)^{\rm T},\; (0,1,0,\cdots,0)^{\rm T},$ and so on, where T denotes
the transpose operation. Because of being multiplied by the successive TMs, the
directions of the initial vectors change. The Lyapunov exponents are then the
logarithm of the trace of the matrix, ${\bf V}=[{\bf T}
({\bf T})^{\rm T}]^{1/2n}$, where, ${\bf T}={\bf T}_n{\bf T}_{n-1}\cdots
{\bf T}_1$. Then, the direction that corresponds to the largest Lyapunov is the
direction of the eigenvector that corresponds to the largest eigenvalue of the
matrix {\bf V}.

However, after a few steps, the information about all the Lyapunov exponents
but the largest one will be lost in the numerical noise. To avoid this
difficulty we implemented the Gram-Schmidt (GS) orthogonalization after
every two steps of the TM iterations. The number of steps after which the GS
orthogonalization should be applied depends on the model, and may actually be
estimated.$^{22,23}$ For example, in the computations for the Anderson model
the GS orthogonalization is applied after every 10 steps.

\begin{figure}[b]
\epsfxsize7truecm \epsffile{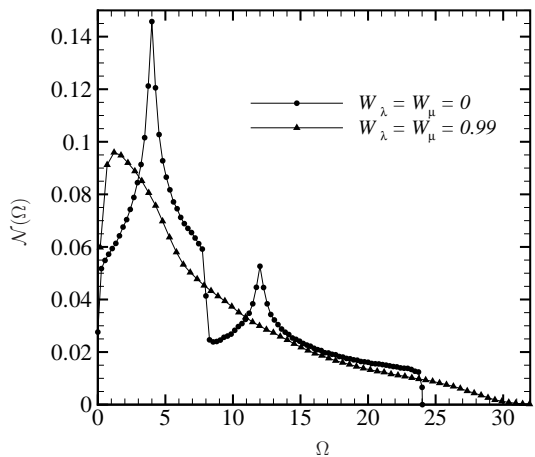} \caption{Density of states
${\cal N}(\Omega)$ of the 2D model, calculated with a computational
grid of size $1000\times 1000$.}
\end{figure}

\begin{center}
{\bf IV. DENSITY OF STATES}
\end{center}

As mentioned above, a useful quantity for characterizing propagation of elastic
waves in a disordered solid is the vibrational density of states (DOS) ${\cal
N}(\omega)$. The usefullness of ${\cal N}(\omega)$ is due to its dependence on
the strength of the disorder and, therefore, it is an important characteristics
of heterogeneous media. To compute the DOS we used the forced-oscillator (FO)
method,$^{5,24,25}$ which made it is possible to use computational grids of
size, $10^3\times 10^3$ in 2D and, $10^2\times 10^2\times 10^2$ in 3D. The FO
method is based on the principle that a complex mechanical system, driven by a
periodic external force of frequency $\Omega$, responds with a large amplitude
in those eigenmodes that are close to $\Omega$.

Let us point out that, although computing the DOS is not directly crucial for
the search for a mobility edge, it is, nevertheless, a very useful quantity to
calculate. For example, the DOS helps one to identify the allowed range of the
frequencies (see Figs. 2 and 3 below), with the help of which the computations
are carried out more efficiently for the mobility edge search in the frequency
range.

\begin{center}
{\bf V. RESULTS AND DISCUSSIONS}
\end{center}

In what follows, we describe the results of the numerical simulations, and
discuss their implications.

\begin{figure}[t]
\epsfxsize7truecm \epsffile{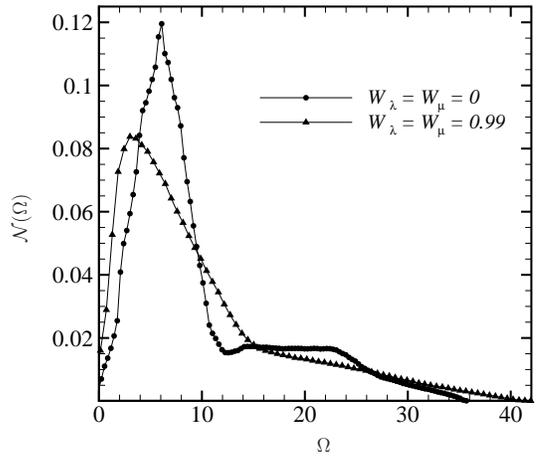} \caption{Density of states
${\cal N}(\Omega)$ of the 3D model, calculated with a computational
grid of size $100\times 100\times 100$. The curves are guide to the
eye.}
\end{figure}

\begin{center}
{\bf A. Density of states}
\end{center}

For an elastic medium in which wave propagation is described by Eq. (1), all
the eigenvalues of the matrix {\bf H} are positive. As a result, the system
undergoes an asymmetric broadening of the band of the allowed frequencies
(energies). In Fig. 2 we present the DOS ${\cal N}(\Omega)$ for the 2D ordered,
as well as random, media. There are two peaks corresponding to two branches of
the characteristic equation, i.e., the transverse and longitudinal modes. The
two modes propagate independently in 2D and, therefore, each branch has the
same DOS as that of the 2D Anderson model. As is well known, without disorder
the DOS has cusps in 2D, containing some special points that are usually called
the von Hove singularities, at which the DOS is nondifferentiable. As Fig. 2
indicates, by adding randomness to the Lam\'e coefficients of the medium, the
von Hove singularities disappear and a band tail appears in the upper band
edge.

\begin{table*}
\setlength{\tabcolsep}{18pt} \centering
\parbox{12.5cm}{\caption{ The rescaled Lyapunov exponents $M\gamma_m$ (inverse of
the rescaled localization length) of the 2D model and the
corresponding errors. $n$ is the index number of $\gamma_m$. The
results are for $M=6$, $L=10^5$, $\Omega=1$, $W_\lambda=W_\mu=0.99$,
and $\lambda_0=\mu_0=1.0$. The Gram-Schmidt orthogonalization was
implemented after every two steps.}}

\begin{tabular}{cccccc}
\hline
$n$ & $M\gamma_m$ & error & $n$ & $M\gamma_m$ & error\\
\hline 1 & 15.421216 & 0.018&13 & -0.146944 & 0.007 \\  2 &
13.044478 & 0.015&14 & -0.384629 & 0.007 \\  3 & 10.675876 &
0.012&15 & -0.769397 & 0.008 \\  4 &  8.019660 & 0.010&16 &
-1.404398 & 0.009 \\  5 &  6.411892 & 0.009&17 & -2.559604 &
0.010 \\  6 &  5.138542 & 0.008&18 & -3.831230 & 0.010 \\
 7 &  3.831297 & 0.008&19 & -5.138530 & 0.009 \\  8 & 2.559524
& 0.008&20 & -6.411741 & 0.010 \\  9 &  1.404456 & 0.007&21 &
-8.019935 & 0.012 \\  10 &  0.769496 & 0.006&22 & -10.675919 & 0.014
\\  11 &  0.384616 & 0.006&23 & -13.044517
& 0.016 \\  12 &  0.146903 & 0.006&24 & -15.421134 & 0.018 \\
\hline
\end{tabular}
\end{table*}

Figure 3 presents the DOS ${\cal N}(\Omega)$ for the corresponding 3D media.
The results seem similar to those for the 2D media, although the cusps do not
appear as sharp. Once again, disorder in the form of the spatial distributions
of the Lam\'e coefficients make the DOS smooth. We show in the next section,
however, that propagation of elastic waves in the the 2D and 3D media is
different, if we study its localization properties.

\begin{center}
{\bf B. Lyapunov exponent and localization length}
\end{center}

The computed Lyapunov exponents $\gamma_m$ are presented in Table I for
$\Omega=1$. They occur in pairs, $(\gamma,-\gamma)$, which indicate the
symplectic symmetry of the TMs. We found that after every two steps the GS
orthogonalization must be implemented. The estimated errors shown were
computed as follows. After each GS orthogonalization the length $d_\alpha$ is
computed for normalizing the evolving vector ${\bf Z}_n$ in Eq. (5), which
then results in a sequence $\{d_\alpha\}$. The average of the logarithm of
such lengths yields the LE,
\begin{equation}
\gamma=\frac{1}{mp}\sum_{\alpha=1}^N\ln(d_\alpha),
\end{equation}
after normalizing the vectors $p$ times, with $m$ being the steps of
GS orthogonalization (here $m=2$). Moreover, the error in estimating
$\gamma$ (see Table I) is given by,$^{26}$
\begin{equation}
\frac{\Delta\gamma}{\gamma}=\frac{1}{\sqrt{p}}\frac{\sqrt{\langle
(\ln d_\alpha)^2\rangle-\langle\ln d_\alpha\rangle^2}}{\langle\ln
d_\alpha\rangle}\;,
\end{equation}
where the brackets indicate averaging over the sequence of $\{d_\alpha\}$.
$\gamma$ is a self-averaged quantity,$^{26}$ and the error of its estimates
approaches zero as $p$ increases. Note that, not all the Lyapunov exponents
are independent; we only need to compute the first $N$ of them. The smallest
positive Lyapunov exponent, $\gamma_m$, corresponds to the localization length
that we wish to determine and study.

\begin{figure}[b]
\epsfxsize7truecm \epsffile{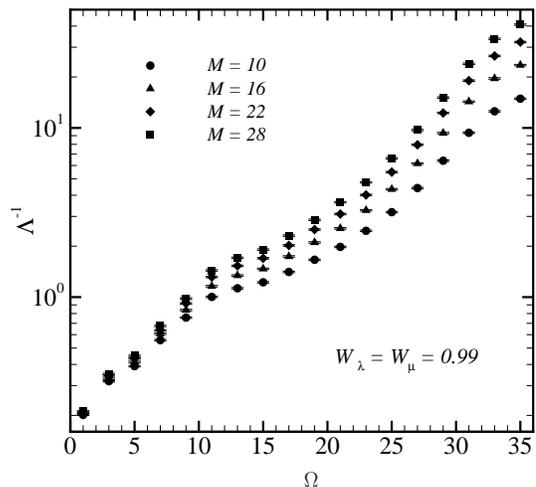} \caption{Inverse of the scaled
localization length $\Lambda$ of the 2D model, obtained with an
accuracy of 0.1\%.}
\end{figure}

Figure 4 presents the inverse of the scaled localization length, $\Lambda=
(M\gamma_m)^{-1}$, in the 2D media as a function of $\Omega=\omega^2$ for the
disorder parameters, $W_\lambda=W_\mu=0.99$ and $\lambda_0=\mu_0=1.0$, and
several $M$, the width of the strips used in the TM computations. The
calculations were carried out with an accuracy of 0.1\%. At first glance, it
appears that $\Lambda$ decreases by increasing the width, implying that it
vanishes in the thermodynamic limit and, therefore, all the states are
localized in the 2D disordered media that we studied, up to the frequencies
that were considered.

However, a closer inspection of Fig. 4 indicates an intriguing possibility.
It appears that for $\Omega>10$, $\Lambda^{-1}$ exhibits an $M-$dependence,
whereas for $\Omega<10$ the dependence on $M$ is weak, if it exists at all.
The difference indicates that the possibility of the existence of a mobility
edge cannot be completely ruled out in such 2D media. If mobility edge does
exist, it would then be consistent with the prediction of the RG
calculations$^{12}$ for the 2D media, which did predict the existence of such
a mobility edge in 2D. We shall come back to this point shortly.$^{28}$

\begin{figure}[t]
\epsfxsize7truecm \epsffile{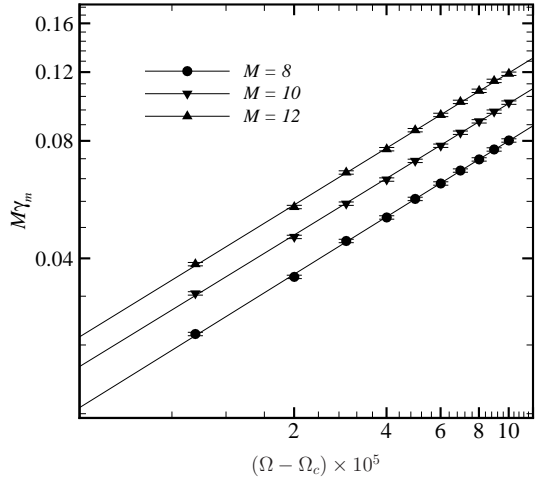} \caption{Inverse of the scaled
localization length $\Lambda$ of the $M\times L$ strips with
$W_\mu=0$ and $W_\lambda=0.99$. The results were obtained with an
accuracy of 0.1\%.}
\end{figure}

Next, we consider a special case in which the shear modulus $\mu$ is constant,
but the bulk modulus $\lambda$ is distributed randomly. Such a limiting case is
of interest, because a 2D medium of this type does have a band of extended
states. It is straightforward to show that, for a constant $\mu$, Eqs. (A1) and
(A2) have a solution that propagates without any scattering which, in fact,
represents plane waves with transverse polarization, for which the dispersion
relation is given by
\begin{equation}
\Omega=4-2\cos(k_x)-2\cos(k_y)\;,
\end{equation}
where ${\bf k}=(k_x,k_y)$ is the wave vector. The dispersion relation (6) has
a frequency band in the range, $0<\Omega<8$, in which we find a zero Lyapunov
exponent, implying infinite localization length and, therefore, extended
states. Moreover, there is a mobility edge at $\Omega_c=8$, and as $\Omega\to
\Omega_c^+$, one has
\begin{equation}
\gamma_m\propto (\Omega-\Omega_c)^{\nu_T}\;.
\end{equation}
Figure 5 presents the frequency-dependence of the rescaled Lyapunov exponent,
$M\gamma_m$, near $\Omega_c=8$ for several strip widths $M$. The data are
well-fitted by the power law (7), yielding the estimate, $\nu_T\simeq 0.496
\pm 0.003$, independent of the width $M$.

\begin{figure}[t]
\epsfxsize7truecm \epsffile{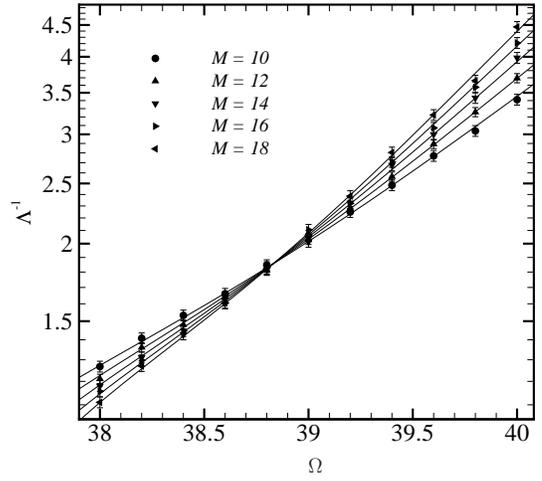} \caption{Inverse of the scaled
localization length $\Lambda$ of the 3D model for
$W_\mu=W_\lambda=0.99$, obtained with an accuracy of 2\%. Solid
curves represent the fit of the data to Eq. (10).}
\end{figure}

In 3D media, however, the Lyapunov exponent behaves differently. Figure 6
displays the results for the inverse of the (rescaled) localization length
$\Lambda^{-1}$ for the same disorder parameters as those in the 2D media. The
results for all values of the width $M$ of the 3D bars (used in the TM
computations) intersect one another at a particular critical frequency
$\Omega_c$. For $\Omega<\Omega_c$ the rescaled Lyapunov exponent decreases
with increasing $M$, i.e., one has extended states.

To obtain better understanding of the localization properties, we analyze the
numerical data for the localization length. The inverse of the (rescaled)
localization length, $\Lambda,^{-1}$ is a function of a single scaling
variable, and is expressed as
\begin{equation}
\Lambda^{-1}=F(\chi M^{1/\nu})= \sum_{i=0}^n a_i\chi^i M^{i/\nu}\;,
\end{equation}
where $\nu$ is the localization length exponent, $\Lambda \sim
(\Omega-\Omega_c),^{-\nu}$ and $\chi$ is the scaling variable. $a_0$ represents
the critical value of $\Lambda_c^{-1}$. The absolute scale of the argument in
Eq. (10) is arbitrary; we fix the coefficients by setting $a_1=1$. The scaling
variable $\chi$ is then expanded as a function of the reduced frequency
$\Omega_r$,
\begin{eqnarray}
\chi & = & \sum_{i=1}^m b_i \Omega_r^i\;,\\
\Omega_r & = & (\Omega-\Omega_c)/\Omega_c\;.
\end{eqnarray}
For a large enough system, it is not necessary to keep the higher order terms
of Eqs. (10) and (11) in the critical region near $\Omega_c$. Here, however,
due to the relatively small sizes of the 3D bars that we used in the TM
simulations, we need some of the leading order terms in order to obtain
accurate fit of the data. If the number of terms in the expansions (10) and
(11) are, respectively, selected to be, $n=3$ and $m=2$, we obtain an accurate
fit of the data with
\begin{equation}
\nu\simeq 1.89\pm 0.17\;,\;\;\;\Lambda_c^{-1}\simeq 1.84\pm 0.06\;,\;\;\;
\Omega_c \simeq 38.82\pm 0.06\;,
\end{equation}
where the estimated errors are with 95\% confidence.

The above estimate of the critical exponent $\nu$ is different from the
estimate$^{29}$ of the corresponding exponent for the Anderson model of
electron localization, $\nu\simeq 1.57\pm 0.01$. It is also much larger than
$\nu=3/2$, which was recently derived$^{30}$ based on a semiclassical theory
for the 3D Anderson model of electron localization. The important implication
of the difference is that, the localization-delocalization transition for
elastic waves in the 3D disordered media that we study belongs to a
universality class different from that of the Anderson model. The difference is
presumably related to the different symmetries of the underlying Hamiltonians
for the two phenomena. We note, however, that, due to the relatively large
estimated errors of $\nu$ for the elastic waves, we cannot completely rule out
the possibility that the two models belong to the same universality class.

Similar to the 2D media, the special limit in which the shear
modulus $\mu$ is constant, but the bulk modulus $\lambda$ varies
spatially, may also be studied in 3D media. In this case the
transverse plane waves have a dispersion relation given by
\begin{equation}
\Omega=6-2\cos(k_x)-2\cos(k_y)-2\cos(k_z)\;.
\end{equation}
Therefore, the frequency band of such waves is the interval $0<\Omega<12$.
Figure 7 presents frequency-dependence of the rescaled minimum Lyapunov
exponent, $\gamma_m$. The results indicate that, near the mobility edge
$\Omega_c=12$, $\gamma_m$ follows the same type of power law as in the 2D
media, with an exponent, $\nu_T\simeq 0.481\pm 0.004$, very close to that
estimated for the 2D media and roughly equal to $1/2$.

\begin{center}
{\bf C. Statistics of energy levels}
\end{center}

\begin{figure}[t]
\epsfxsize7truecm \epsffile{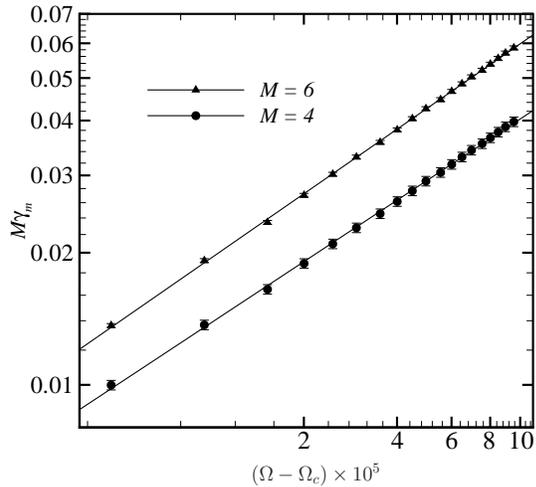} \caption{Inverse of the scaled
localization length $\Lambda$ of the $M\times M\times L$ bars
with$W_\mu=0$ and $W_\lambda=0.99$. The results were obtained with
an accuracy of 0.1\%.}
\end{figure}

\begin{figure}[t]
\epsfxsize7truecm \epsffile{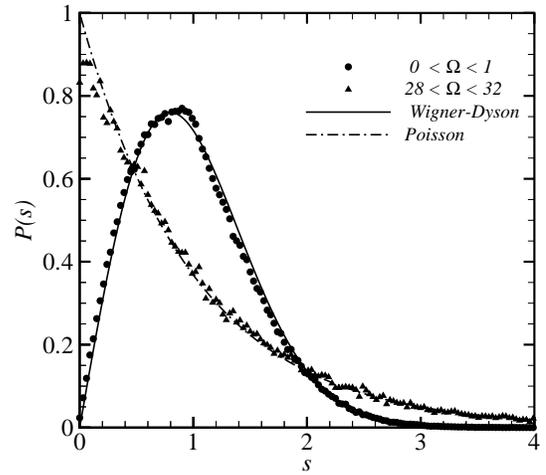} \caption{Distribution of the
level spacings for $10^3$ realizations of 2D media of size $50^2$
for two frequency intervals. The results for the high frequency
interval follows the Poisson statistics, while the results for the
low-frequency interval follows the statistics of the Gaussian
orthonormal ensemble.}
\end{figure}

\begin{figure}[b]
\epsfxsize7truecm \epsffile{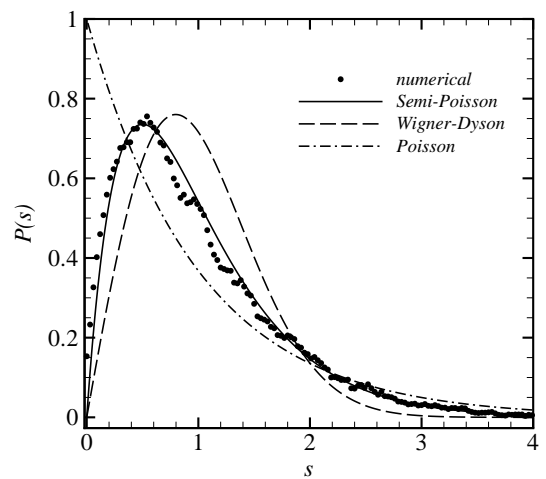} \caption{Distribution $P(s)$
of the level spacings for 120 realizations of the 3D media of size
$30^3$ for the frequency interval $36<\Omega<40$, near the critical
frequency $\Omega_c$. $P(s)$ is nearly semi-Poisson.}
\end{figure}

An important aspect of the symmetry of any Hamiltonian is its level-spacing
statistics, which have been studied extensively by the theory of random
matrices. The statistics of the level spacings $s$ for the elastic waves
indicate that, in the localized regime where the localization length is small
compared to the medium's linear size, the levels are uncorrelated and,
therefore, they follow a Poisson distribution. Figure 8 presents such
energy-level statistics for the 2D media for two frequency ranges. The paramter
$s$ is defined by
\begin{displaymath}
s=\frac{\Omega_{n+1}-\Omega_n}{\langle \Omega_{n+1}-\Omega_n\rangle}\;.
\end{displaymath}
The results, shown in Fig. 8, were obtained by exact diagonalization
of 1000 realizations of the disorder using $50\times 50$
computational grids. In the interval $28<\Omega< 32$ the statistics
do follow the Poisson distribution. For low frequencies in the
interval $0<\Omega<1$, however, the data are well-fitted by the
Wigner-Dyson distribution. Indeed, the Wigner-Dyson and Poisson
distributions represent the statistics for certain limits that are
attained in the thermodynamic limit. In fact, the Wigner-Dyson
distribution, which describes the statistics of the level spacings
in the low frequency limit, will approach the Poisson distribution
by increasing the system size,$^{31}$ if there is no true mobility
edge in two dimensions.

At the localization-delocalization transition point in 3D, the distribution
function of the level spacings is independent of the system's size, and is
represented by the semi-Poisson distribution,$^{32}$ $P(s)=4s\exp(-2s)$.
Figure 9 presents the distribution function of the level spacings in the 3D
model for the frequencies near the critical frequency $\Omega_c$, obtained by
using 120 realizations of the disorder in $30\times 30\times 30$ computational
grids. The results indicate clearly that the distribution is essentially
semi-Poisson.

\begin{center}
{\bf D. Comparison with the predictions of the dynamic renormalization}
\end{center}

In a previous paper$^{12}$ we studied localization of elastic waves in 2D
heterogeneous solids with randomly distributed Lam\'e coefficients, as well as
those with long-range correlations with a power-law correlation function,
characterized by an exponent $\rho$. The Martin-Siggia-Rose method$^{13}$ was
used, and the one-loop RG equations for the the coupling constants were derived
in the limit of low frequencies (long wavelengths). We found, in particular,
that for $\rho<1$ there is a region of the coupling constants space in which
the RG flows are toward the Gaussian fixed point, implying that the disorder
is irrelevant and, therefore, the waves are delocalized. In the rest of the
disorder space the elastic waves were found to be localized.

The numerical results for the 2D media presented in the present paper, when
the two Lam\'e coefficients are spatially distributed, do not seem to indicate
the existence of any extended states. If this is true, then the discrepancy
between the numerical simulations and the RG predictions may be due to the fact
that, the RG results are valid only in the low-frequency limit, and the
smallest frequencies that we considered in the present paper may be larger than
those for which the RG results are valid.

At the same time, as discussed above, the $M-$dependence of the inverse
(rescaled) localization length $\Lambda^{-1}$ for a range of frequencies
$\Omega>10$, and its absence for $\Omega<10$, open up the possibility that,
consistent with the RG predictions, a mobility edge does exist in the 2D media
that we study. In that case, we would have further evidence that question of
localization of elastic waves in heterogeneous media is fundamentally different
from that of Anderson localization of electrons. This is clearly an issue that
deserves further study. Work in this direction is in progress.

\begin{center}
{\bf VI. SUMMARY}
\end{center}

We studied the localization properties of elastic waves in a disordered
elastic medium in both two and three dimensions. We found that if the
disorder, in the form of spatially-varying Lam\'e coefficients, is broadly
distributed, it may lead to the localization of all the states in the 2D media,
although there is some evidence that a mobility edge might exist in such
media. The same disorder strength cannot, however, localize all the states in
the 3D media. There is a mobility edge in 3D, near which the (rescaled)
localization length follows a power law, $\Lambda\sim(\Omega-\Omega_c)^{-\nu}$.
Using extensive numerical simulations and a scaling analysis, the estimated
$\nu$ was found to be different from that of the Anderson model of electron
localization. The statistics of the energy levels indicated, however, that in
the extended regime it is generally the Gaussian orthogonal ensemble
statistics$^{11}$ that govern the energy levels. In the limit in which the
shear modulus is constant but the bulk modulus varies spatially, there is a
mobility edge even in 2D, for which the associated critical exponent $\nu_T$
for the power-law behavior of the localization length near the edge was also
computed. We found for both the 2D and 3D media that, $\nu_T\simeq 1/2$.

\begin{center}
{\bf APPENDIX: THE FINITE-DIFFERENCE EQUATIONS}
\end{center}

\setcounter{equation}{0}
\renewcommand{\theequation}{A.{\arabic{equation}}}

We list the discretized equations that govern the displacements of the grid
points during propagation of elastic waves in the 2D and 3D computational
grids.

\begin{center}
{\bf Two-dimensional media}
\end{center}
\begin{eqnarray}
-m\omega^2 u & = & D_x[(\lambda+2\mu)D_x u] + D_y(\mu D_y u) + D_y
(\mu D_x w) \nonumber\\&&+ D_x(\lambda D_y w)|_{i,j}\;,\\
-m\omega^2 w & = & D_y[(\lambda+2\mu)D_y w] + D_x(\mu D_x w) + D_x
(\mu D_y u) \nonumber\\&&+ D_y(\lambda D_x
u)|_{i+\frac{1}{2},j+\frac{1}{2}}\;.
\end{eqnarray}
where
\begin{eqnarray}
D_x f(i,j) & = & \frac{1}{h}\left[f(i+\frac{1}{2},j)-f(i-\frac{1}{2},j)\right]
\;,\\
D_y f(i,j) & = & \frac{1}{h}\left[f(i,j+\frac{1}{2})-f(i,j-\frac{1}{2})\right]
\;.
\end{eqnarray}

\begin{center}
{\bf Three-dimensional media}
\end{center}
\begin{eqnarray}
-m\omega^2 u & = & D_x[(\lambda+2\mu)D_x u] + D_y(\mu D_y u) +
D_z(\mu D_z u) \nonumber\\&&+ D_y (\mu D_x w) + D_x(\lambda D_y w) + D_x(\mu D_z v)
\nonumber\\&&+ D_z(\lambda D_x v)|_{i,j,k}\;,\\
-m\omega^2 w & = & D_y[(\lambda+2\mu)D_y w] + D_x(\mu D_x w) +
D_y(\mu D_y w) \nonumber\\&&+ D_x (\mu D_y u) + D_y(\lambda D_x u) + D_y(\mu D_z v)
\nonumber\\&&+ D_z(\lambda D_y v)|_{i+\frac{1}{2},j+\frac{1}{2},k}\;,\\
-m\omega^2 v & = & D_z[(\lambda+2\mu) D_z v] + D_x(\mu D_x v) +
D_y(\mu D_y v) \nonumber\\&&+ D_z (\mu D_x u) + D_x(\lambda D_z u) +
D_z (\mu D_y w) \nonumber\\&&+ D_y(\lambda D_z
w)|_{i+\frac{1}{2},j,k+\frac{1}{2}}\;.
\end{eqnarray}
where
\begin{eqnarray}
D_x f(i,j,k) & = & \frac{1}{h}\left[f(i+\frac{1}{2},j,k)-f(i-\frac{1}{2},j,k)
\right]\nonumber\\\;,\\
D_y f(i,j,k) & = &
\frac{1}{h}\left[f(i,j+\frac{1}{2},k)-f(i,j-\frac{1}{2},k)
\right]\nonumber\\\;,\\
D_z f(i,j,k) & = &
\frac{1}{h}\left[f(i,j,k+\frac{1}{2})-f(i,j,k-\frac{1}{2})
\right]\nonumber\\\;.
\end{eqnarray}
We took $h=1$.

\begin{center}
\line(1,0){230}
\end{center}

\noindent\textsuperscript{*}moe$@$iran.usc.edu

\begin{description}

\item $^1$A. Ishimaru, {\it Wave Propagation and Scattering in Random Media}
(Academic, New York, 1978).

\item $^2$N. Bleistein, J. K. Cohen, and J. W. Stockwell, Jr., {\it Mathematics
of Multidimensional Seismic Imaging, Migration, and Inversion} (Springer, New
York, 2001).

\item $^3$A. V. Granato and K. L\"ucke, in {\it Physical Acoustics}, edited by
W. P. Mason (Academic, New York, 1966), Vol. 4a.

\item $^4$M. R. Rahimi Tabar, M. Sahimi, F. Ghasemi, K. Kaviani, M.
Allamehzadeh, J. Peinke, M. Mokhtari, M. Vesaghi, M. D. Niry, A. Bahraminasab,
S. Tabatabai, S. Fayyazbakhsh, and M. Akbari, in {\it Modelling Critical and
Catastrophic Phenomena in Geoscience}, edited by P. Bhattacharya and B. K.
Chakrabarti (Springer, Berlin, 2006); P. Manshoor, S. Saberi, M. Sahimi, J.
Peinke, A. F. Pacheco, and M. R. Rahimi Tabar (unpublished).

\item $^5$M. Sahimi, {\it Heterogeneous Materials I \& II} (Springer, New
York, 2003).

\item $^6$See, for example, J. B. Keller, Proc. Symp. Appl. Math. {\bf 16},
145 (1964); T. Yamashita, Pure Appl. Geophys. {\bf 132}, 545 (1990).

\item $^7$J. E. Gubernatis, E. Domany, and J. A. Krumhansl, J. Appl. Phys.
{\bf 48}, 2804 (1977).

\item $^8$E. Larose, L. Margerin, B. A. van Tiggelen, and M. Campillo, Phys.
Rev. Lett. {\bf 93}, 048501 (2004).

\item $^9$P. W. Anderson, Phys. Rev. {\bf 109}, 1492 (1958).

\item $^{10}$E. P. Wigner, Ann. Math. {\bf 53}, 36 (1951); F. J. Dyson, J.
Math. Phys. {\bf 3}, 140 (1962).

\item $^{11}$M. L. Mehta, {\it Random Matrices} (Academic, Boston, 1991).

\item $^{12}$R. Sepehrinia, A. Bahraminasab, M. Sahimi, and M. R. Rahimi Tabar,
Phys. Rev. B {\bf 77}, 014203 (2008).

\item $^{13}$P. C. Martin, E. G. Siggia, and H. A. Rose, Phys. Rev. A {\bf 8},
423 (1973).

\item $^{14}$P. H. Song and D. S. Kim, Phys. Rev. B {\bf 54}, R2288 (1996).

\item $^{15}$Y. Akita and T. Ohtsuki, J. Phys. Soc. Jpn. {\bf 67}, 2954 (1998).

\item $^{16}$J. J. Ludlam, S. N. Taraskin, and S. R. Elliot, Phys. Rev. B
{\bf 67}, 132203 (2003).

\item $^{17}$T.-K. Hong and B. L. N. Kennett, Geophys. J. Int. {\bf 150}, 610
(2002); {\it ibid.} {\bf 154}, 483 (2003).

\item $^{18}$Y. Q. Zeng and Q. H. Liu, J. Acous. Soc. Amer. {\bf 109}, 2571
(2001).

\item $^{19}$T. Bohlen, Comput. Geosci. {\bf 28}, 887 (2002).

\item $^{20}$H. A. Friis, T. A. Johansen, M. Haveraaen, H. Munthe-Kaas, and A.
Drottning, Appl. Numer. Math. {\bf 39}, 151 (2001).

\item $^{21}$M. Sahimi and S. M. Vaez Allaei, Comput. Sci. Eng. {\bf 10}
(No. 3), 66 (2008).

\item $^{22}$A. MacKinnon and B. Kramer, Z. Phys. B {\bf 53}, 1 (1983).

\item $^{23}$J. L. Pichard and G. Sarma, J. Phys. C {\bf 14} L127 (1981).

\item $^{24}$M. L. Williams and H. J. Maris, Phys. Rev. B {\bf 31}, 4508
(1985).

\item $^{25}$K. Yakubo and T. Nakayama, Phys. Rev. B {\bf 40}, 517 (1989);
K. Yakubo, T. Nakayama, and H. J. Maris, J. Phys. Soc. Jpn. {\bf 60}, 3249
(1990); K. Yakubo, K. Takasugi, and T. Nakayama, {\it ibid.} {\bf 59}, 1909
(1990); T. Nakayama and K. Yakubo, Phys. Rep. {\bf 349}, 239 (2001).

\item $^{26}$R. Sepehrinia, M. D. Niry, N. Bozorg, M. R. Rahimi Tabar, and M.
Sahimi, Phys. Rev. B {\bf 77}, 104202 (2008).

\item $^{27}$P. Markos, arXiv:cond-mat/0609580v1

\item $^{28}$The subject remains controversial. See, for example, P. Markos
and C. M. Soukoulis, Phys. Rev. B {\bf 71}, 054201 (2005).

\item $^{29}$K. Slevin and T. Ohtsuki, Phys. Rev. Lett. {\bf 82}, 382 (1999).

\item $^{30}$A. M. Garc\'ia-Garc\'ia, Phys. Rev. Lett. {\bf 100}, 076404
(2008).

\item $^{31}$I. Kh. Zharekeshev, M. Batsch, and B. Kramer, Europhys. Lett.
{\bf 34}, 587 (1996).

\item $^{32}$D. Braun, G. Montambaux, and M. Pascaud, Phys. Rev. Lett. {\bf
81}, 1062 (1998); E. B. Bogomolny, U. Gerland, and C. Schmit, Phys. Rev. E.
{\bf 59}, 1315 (1999); S. N. Evangelou, J. Phys. A {\bf 38}, 363 (2005).

\end{description}

\end{document}